\documentclass{article}

\usepackage{PRIMEarxiv}

\usepackage[utf8]{inputenc} 
\usepackage[T1]{fontenc}    
\usepackage{hyperref}       
\usepackage{url}            
\usepackage{booktabs}       
\usepackage{amsfonts}       
\usepackage{nicefrac}       
\usepackage{microtype}      
\usepackage{lipsum}
\usepackage{float}
\usepackage[numbers]{natbib} 
\usepackage{fancyhdr}       
\usepackage{graphicx}       
\graphicspath{{media/}}     

\title{Quantum-Enhanced Secure Approval Voting Protocol
}

\author{
  Saiyam Sakhuja \\
  Department of Physics \\
  National Institute of Technology Tiruchirappalli \\
  Tiruchirappalli, Tamil Nadu, India\\
  \texttt{sakhujasaiyam@gmail.com} \\
   \And
  S. Balakrishnan \\
  Department of Physics, School of Advanced Sciences \\
  Vellore Institute of Technology \\
  Vellore, Tamil Nadu, India\\
  \texttt{physicsbalki@gmail.com} \\
}

\begin{document}
\maketitle

\begin{abstract}
In a world where elections touch every aspect of society, the need for secure voting is paramount. Traditional safeguards, based on classical cryptography, rely on complex math problems like factoring large numbers. However, quantum computing is changing the game. Recent advances in quantum technology suggest that classical cryptographic methods may not be as secure as we thought. This paper introduces a quantum voting protocol, a blend of quantum principles (entanglement and superposition), blockchain technology, and digital signatures, all powered by $\log_2{n}$ qubits, and designed for approval voting with $n$ candidates. The result is a symphony of security features - binding, anonymity, non-reusability, verifiability, eligibility, and fairness - that chart a new course for voting security. The real world beckons, as we tested this protocol on IBM quantum hardware, achieving impressively low error rates of just 1.17\% in a four-candidate election.
\end{abstract}


\section{Introduction} 
The act of casting a vote is widely acknowledged as a potent mechanism through which individuals can express their opinions on a particular subject matter. In principle, the preferences of voters and the selected voting methodology have the capacity to impact the final outcome of the vote \cite{ku1999secure}. Elections serve as a fundamental process for decision-making across diverse professional and personal contexts. Elections take various forms and find extensive applications, ranging from the selection of student representatives within educational institutions to the appointment of chairpersons in corporate settings and the election of national leaders. With the rapid advancement of the information age, electronic elections are progressively replacing traditional paper-based voting systems, aligning more closely with our daily routines and professional engagements. Following Chaum's proposal \cite{chaum1988multiparty} of the first electronic voting protocol in 1981, numerous traditional electronic election protocols have been developed. However, the shift to electronic voting also introduces the potential for adversaries to exert influence or disrupt the voting process more easily, particularly in the presence of minor security vulnerabilities in the system's design \cite{gao2021quantum}. Fujioka underscored the significance of several fundamental security requisites within a secure electronic voting scheme, which encompass eligibility, verifiability, accuracy, and the confidentiality of the ballot \cite{fujioka1993practical}. \\
Unfortunately, the security of classical cryptography hinges on unverified assumptions pertaining to the computational complexity of specific mathematical functions, notably the challenge of factoring large integers. Recent developments in quantum computation \cite{shor1999polynomial} raise the possibility that quantum computers could substantially expedite the process of factoring large numbers in contrast to classical computers. Consequently, classical cryptographic techniques are presently susceptible to potential attacks based on quantum computing capabilities \cite{xue2017simple}. Quantum computing, characterized by its distinctive attributes and formidable computational capabilities, represents a promising pathway for the fundamental transformation of the voting process. Quantum voting protocols, firmly rooted in the principles of quantum mechanics, offer a viable approach to tackle the longstanding challenges confronting traditional voting systems. The concept of quantum voting has elicited substantial attention within the research community owing to its capacity to withstand potential threats posed by quantum algorithms \cite{wang2020quantum}. Hillery \cite{hillery2006quantum} introduced a pair of voting methods, specifically distributed voting and traveling voting. Vaccaro  \cite{vaccaro2007quantum} introduced criteria for quantum voting systems. Subsequently, Tian \cite{tian2016voting} put forward a voting approach based on entangled states. Entanglement, a fundamental quantum phenomenon, creates a unique and inseparable connection between quantum particles, impacting various aspects of quantum technologies and cryptography. As the field of research progressed, numerous quantum voting methods were proposed by scholars, for instance, \cite{horoshko2011quantum, wang2019fault, jiang2020nonlocal, xu2018quantum, jiang2020quantum, xu2022quantum}. \\
Researchers have explored various voting methods, each with its own distinctive approach to candidate evaluation. Range voting \cite{harsanyi1986rational} involves voters assigning points to candidates independently, and the candidate with the highest total points wins. Majority judgment \cite{balinski2011majority}, on the other hand, relies on the candidate with the highest median number of points as the winner. In approval voting \cite{brams1978approval}, voters can approve or disapprove of (thumbs up or thumbs down) one or more candidates, without needing to rank them. The candidate with the most approvals wins the election. Approval voting is notable for its simplicity, which can lead to increased voter participation and clear winner determination. \\
For voting protocols to be considered reliable and practical, they must adhere to a set of desirable properties \cite{sun2019simple}, including:

\begin{enumerate}
  \item Anonymity: Ensuring that only the voter knows the content of their vote.
  \item Binding: Preventing any unauthorized alteration of a ballot after submission.
  \item Non-reusability: Guaranteeing that each voter can cast only one vote.
  \item Verifiability: Allowing every voter to confirm the accurate counting of their ballot.
  \item Eligibility: Restricting the voting process to eligible voters only.
  \item Fairness: Ensuring that no one can access partial ballot tallies before the official tally.
\end{enumerate}

In the present work, we introduce a novel voting protocol designed for multiple candidates utilizing the approval voting method. Harnessing the distinctive qualities of quantum superposition \cite{dirac1981principles}, quantum entanglement \cite{einstein1935can}, blockchain technology \cite{nakamoto2008bitcoin}, and cryptographic signatures, the presented protocol introduces a groundbreaking method for upholding the integrity, security, and confidentiality of the voting procedure. Furthermore, we have implemented and rigorously analyzed this voting protocol using the Qiskit \cite{aleksandrowicz2019qiskit} framework for a scenario involving four candidates. Specifically, we employ the Amplitude Encoding Technique \cite{larose2020robust} for the encoding of votes into quantum states within our Quantum Voting Protocol. This technique, rooted in the principle of superposition, allows for the concurrent representation of multiple quantum states within a single qubit. \\
The paper is structured into six sections. Section 2 introduces the Quantum Voting Protocol tailored for approval voting systems, elucidating its intricate quantum and cryptographic components. Section 3 conducts a thorough security analysis, evaluating the protocol's resilience against various threats. In Section 4, we present the practical Qiskit implementation for four candidates1, while Section 5 offers interpretations and insights derived from its results. The paper culminates in Section 6, where we summarize findings and discuss the transformative potential of quantum technologies in voting systems.

\section{Proposed quantum voting machine}
\label{sec:headings}

In this voting protocol, the participants and their designated roles are defined as follows: Alice assumes the role of the voter, Bob is designated as the tallyman, and Charlie$_i$ represents the scrutineer. It is imperative to underscore that Charlie$_i$ does not signify an individual but rather a distinct group within the voting organizing committee. Furthermore, Charlie$_i$ will assume responsibility for the management of node addition.

\subsection{Initialize}
\begin{enumerate}
    \item Alice wants to cast her vote.
    \item Bob and Alice checks each other’s credentials. If Alice is eligible, then Bob gives her a unique ID number.
    \item Bob and Alice uses a specific hash function \cite{sahu2017review}, which they only knows, to create a hash ID for their unique ID. Bob stores the hash ID in the voting database.
    \item Bob gives secret keys of $N$ length to Alice and Charlie$_i$. K$_{AB}$, K$_{AC}$ to Alice and K$_{AC}$ to Charlie$_i$. Bob generates the secret keys via Quantum Random Number Generator (QRNG) \cite{herrero2017quantum}. Secret keys are randomly generated binary numbers of length $N$.
\end{enumerate}

\subsection{Voting Procedure}

\begin{enumerate}
    \item Alice casts her vote in her $n$ qubits. $n$ is calculated as the smallest integer greater than or equal to the base-2 logarithm of $N$, denoted as $\lceil{\log_2{n}}\rceil$, i.e., $n \geq \log_2{N}$.
    \item Alice encodes her vote in $n$ qubits with the help of Amplitude Encoding Technique. 
    \item Alice entangles her qubits and adds her signature on the qubits. The signature can be any single or multi-qubit gates. This signature is analogous to a handwritten signature on a paper document. This signature ensures added security and personalization.
    \item Alice sends her qubits to Charlie$_i$. Also, Alice encrypts her hash ID by K$_{AC}$. Alice sends the encrypted message (hash ID) to Charlie$_i$. 
    \item Charlie$_i$ decrypts hash ID with K$_{AC}$.
    \item Alice tells the signing procedure to Charlie$_i$ via classical channel. Charlie$_i$ performs the inverse of the signing procedure to remove the signature from the message. Signature ensures Charlie$_i$ that the qubits are untampered and are only sent by Alice.
    \item Bob shares the voting database with Charlie$_i$.
\end{enumerate}


\subsection{Tally Phase}

\begin{enumerate}

    \item Charlie$_i$ uses Grover's Search Algorithm \cite{grover1996fast} to search Alice's hash ID in the database.
    \item If Charlie$_i$ finds Alice's hash ID in the database, then she will add the node in the quantum blockchain. Node will contain (Alice's signing details + Alice's hash ID + time stamp + hash ID of previous node).
    \item After finding the hash ID in the database, Charlie$_i$ will remove that particular hash ID from the database, so that Alice cannot vote again.
    \item After the node is added in the blockchain, Alice can match her hash ID along with signing details. 
    \item If Alice is convinced that her message is reached to Charlie$_i$ without any tampering, then she will encrypt the entanglement details in K$_{AB}$ and send to Bob.
    \item Charlie$_i$ will share the qubits with Bob to execute the tally process. 
    \item Bob will decrypt the entanglement details with K$_{AB}$ then, he will apply inverse operation on the qubits and will receive the original vote message.
    \item Bob can tally the votes under Charlie$_i$'s supervision and announce the winner.

\end{enumerate}

\section{Security Analysis}
Our voting protocol successfully fulfills the following security requirements:
\begin{enumerate}

    \item Anonymity – To protect Alice's identity, her Unique ID undergoes a hash function, ensuring her anonymity.
    \item Binding – Other voters are unable to alter a voter's choice due to the encoding of each vote in qubits, followed by the introduction of security layers through entanglement and signatures.
    \item Non-reusability – Each voter possesses a single hash ID, and after node addition, the hash ID is promptly removed from the database to prevent multiple voting.
    \item Verifiability – Every voter has the ability to verify if their vote has reached the voting authority without tampering and has been successfully cast.
    \item Eligibility – Bob confirms the eligibility of voters, granting a unique ID solely to eligible individuals.
    \item Fairness – To maintain fairness, the vote message is not directly added to the node. Instead, Bob tallies the votes under the supervision of Charlie$_i$, preventing any premature ballot tallying that could compromise fairness.

\end{enumerate}

\section{Explanation with example}

\begin{figure}[h]
    \centering
    \includegraphics{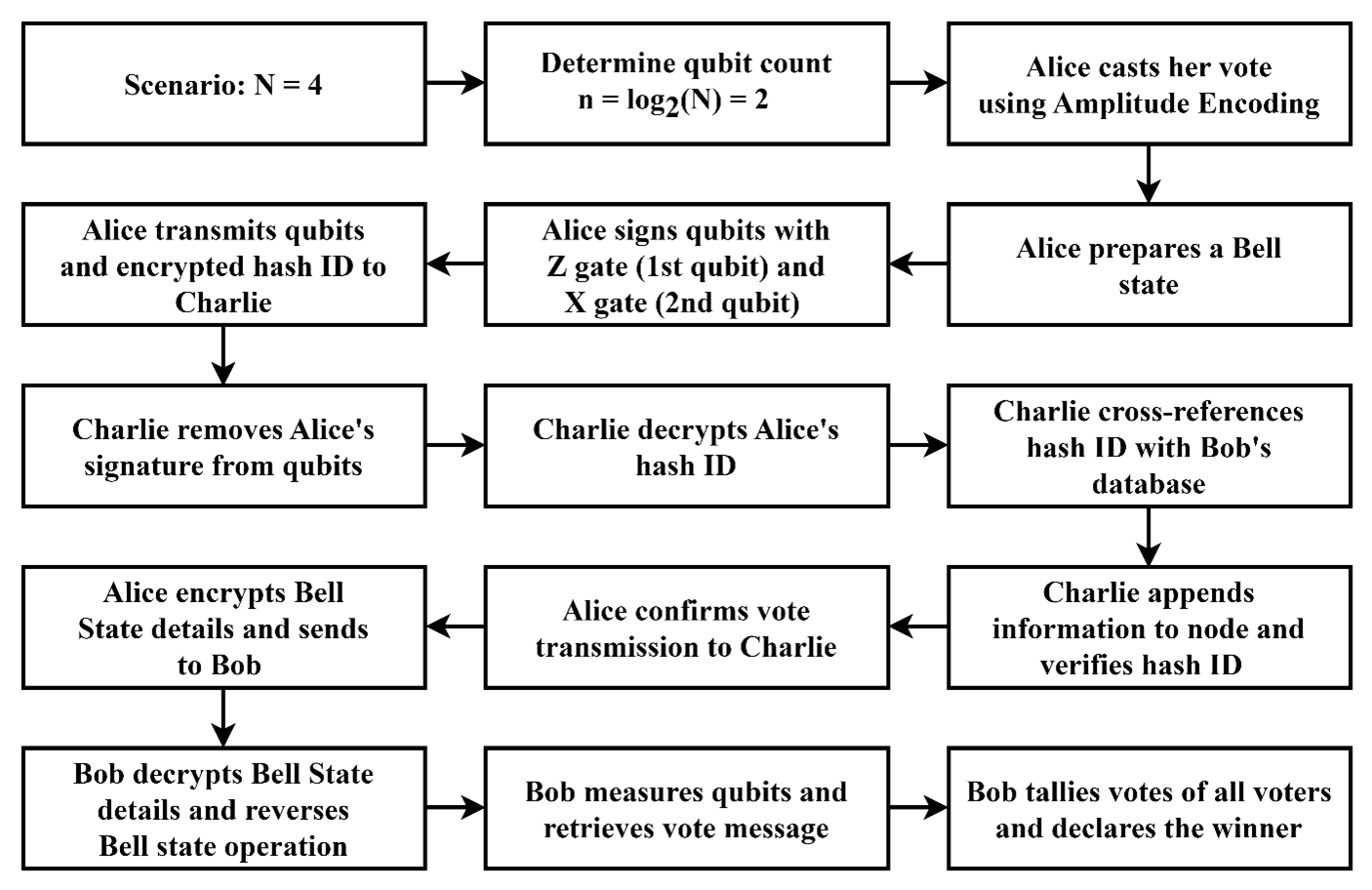}
    \caption{Flowchart depicting the Secure Quantum Voting Protocol for a Scenario Involving Four Candidates.}
    \label{fig:flowchart}
\end{figure}

When considering a scenario involving a total of $N = 4$ candidates, it becomes apparent that the requisite number of qubits, denoted as $n$, can be determined by the equation n = $\log_2{N}$. Consequently, in this case, $n$ equals $\log_2{4}$, which simplifies to $n = 2$. Subsequently, utilizing the Amplitude Encoding Technique, Alice casts her vote in her two qubits. Following the encoding of the vote, Alice proceeds to prepare any one of the Bell states to induce entanglement in her qubits. Additionally, Alice affixes her signature onto the qubits by applying a Z gate to the first qubit and an X gate to the second qubit. The selection of the signature and Bell states is at Alice's discretion. Subsequently, Alice transmits her qubits to Charlie$_i$, along with her encrypted hash ID. Alice communicates her signing details to Charlie$_i$ via a classical channel. Upon reception, Charlie$_i$ applies the appropriate gates (Z gate to the first qubit and an X gate to the second qubit) to the qubits to remove Alice's signature. Charlie$_i$ then decrypts Alice's hash ID and cross-references it with the database provided by Bob. If the hash ID is present in the database, Charlie$_i$ appends a node to the blockchain containing Alice's signing details, hash ID, timestamp, and the hash ID of the previous node. Following verification of her hash ID and personalized signing details, Alice confirms that her vote has securely reached the scrutinizing authority (Charlie$_i$). Alice then encrypts the details of her Bell State and forwards them to Bob. Bob decrypts the Bell State details and applies the appropriate gates to reverse the Bell state operation, thereby retrieving the vote message from Alice. Under the supervision of Charlie$_i$, Bob proceeds to measure the qubits and obtain the actual vote message transmitted by Alice. Fig. \ref{fig:flowchart} shows the flowchart of the entire voting protocol with the help of a scenario of approval voting for four candidates.

\section{Qiskit Implementation for four candidates}
In this section, we present the Qiskit implementation of a Quantum Voting Machine designed for elections with four candidates. Leveraging the power of quantum computing, this implementation offers a novel approach to secure and verifiable voting systems.

\subsection{Implementation Details}
The Quantum Voting Protocol is implemented using the Qiskit framework, a powerful tool for quantum computing development. This implementation focuses on demonstrating the application of quantum principles in the context of voting. \\
When contemplating the scenario with a total of $N = 4$ candidates, it becomes evident that the requisite number of qubits, denoted as $n$, can be determined by the equation $n = \log_2{N}$. Consequently, in this case, n equals $\log_2{4}$, which simplifies to $n = 2$.

In the context of Approval Voting, let us consider Alice's voting preference for a scenario involving four candidates. In her vote, she expresses approval for Candidates 1, 2, and 4, while she disapproves of Candidate 3. This translates to her vote being represented as '1101,' where ‘1’ signifies approval, and '0' signifies disapproval. Utilizing the Amplitude Encoding technique, candidates can be represented by basis as shown in Table \ref{tab:candidates_basis}. The basis corresponding to disapproved candidates, here, Candidate 3 corresponds to $|10\rangle$, have zero (0) as coefficient, while the approved candidates, here, Candidate 1, Candidate 2 and Candidate 4, have non-zero coefficient. Notice, the factor $\frac{1}{\sqrt{3}}$ is normalization factor.

\begin{table}[h]
    \centering    
    \begin{tabular}{|c|c|}  
        \hline
        \textbf{Candidates} & \textbf{Corresponding Basis State} \\  
        \hline
        Candidate 1 & $|00\rangle$ \\  
        \hline
        Candidate 2 & $|01\rangle$ \\  
        \hline
        Candidate 3 & $|10\rangle$ \\  
        \hline
        Candidate 4 & $|11\rangle$ \\  
        \hline
    \end{tabular}
    \caption{Four candidates are represented by four basis states}
    \label{tab:candidates_basis}
\end{table}
Alice's vote can be transformed into a qubit state that would resemble Eq. \ref{eq1}. 

\begin{equation}
\frac{1}{\sqrt{3}} \left( |00\rangle + |01\rangle + |11\rangle \right)
\label{eq1}
\end{equation}

The steps within the Initialization section will remain consistent when using inputs $N=4$ and $n=2$. In the Voting Phase, all the steps will proceed similarly, with the exception of step 2, where Alice may choose to entangle her qubits in any Bell state of her preference. Steps in the Tally Phase section will also remain consistent.

\section{Discussion}
\begin{figure}[!htbp]
  \centering
  \begin{tabular}{@{}c@{\;}c@{}}  
    \parbox{8cm}{\centering \includegraphics[width=8cm]{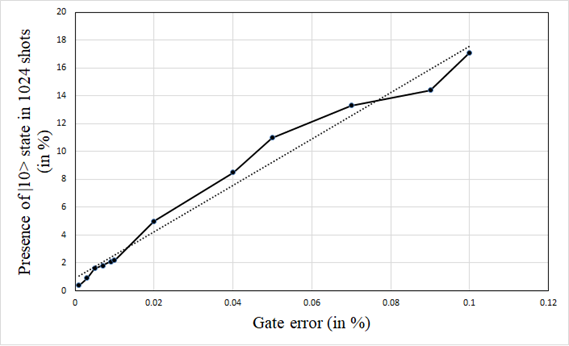}} &  
    \parbox{8cm}{\centering \includegraphics[width=8cm]{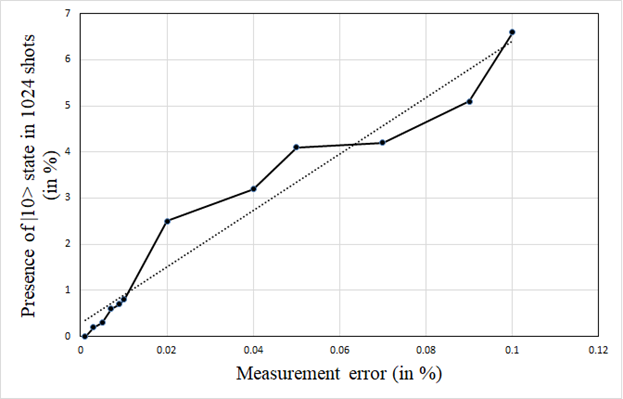}} \\  
  \end{tabular}
  \caption{Variation in Counts (in \%) of $|10\rangle$ basis at variable Gate Error and variable Measurement Error respectively}  
  \label{fig:gate_measurement}
\end{figure}

\begin{figure}[!htbp]
  \centering
  \begin{tabular}{@{}c@{\;}c@{}}  
    \parbox{8cm}{\centering \includegraphics[width=8cm]{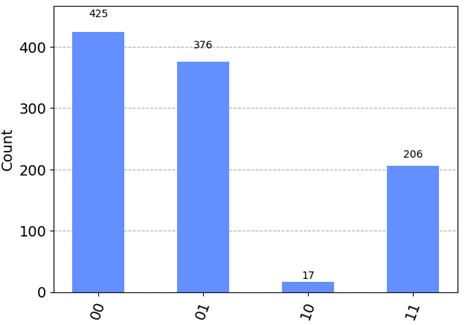}} &  
    \parbox{8cm}{\centering \includegraphics[width=8cm]{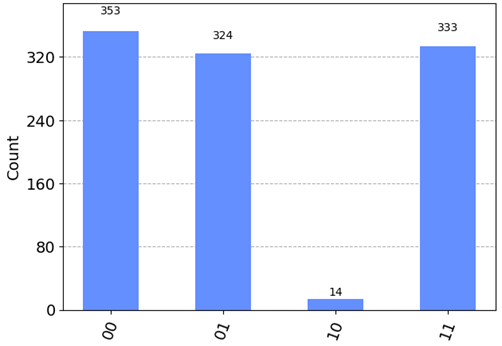}} \\  
  \end{tabular}
  \caption{Results from ibm\_nairobi and ibm\_lagos for four candidates where candidates 1,2, and 4 are approved by the voter and candidate 3 is disapproved.}  
  \label{fig:nairobi_lagos}
\end{figure}

The Qiskit code for implementing the Quantum Voting Protocol for four candidates is executed in a simulated noisy environment with varying error probabilities of gate error and measurement error. Additionally, the code is also executed on two IBM Quantum machines, namely, ibm\_nairobi and ibm\_perth, both of which are 7-qubit devices.
Within the observed error probability ranges, several trends and patterns come to the forefront. Firstly, gate errors exhibit greater prominence when compared to measurement errors for a given probability range, revealing their dominant influence. Notably, when focusing on the counts of the $|10\rangle$ (noise) state among 1024 measurements, a linear correlation with error percentage emerges, as illustrated in Fig. \ref{fig:gate_measurement}. Further exploration within the error probability range of 0.1\% to 1\% exposes a clear linear trend in one of the measurements, while gate errors fluctuate. However, gate errors reclaim their primary position within the 1\% to 10\% error probability range. Additionally, under constant measurement error probability percentages, it becomes evident that errors originating from the $|00\rangle$ and $|11\rangle$ states exert a more substantial impact compared to the $|01\rangle$ state. This phenomenon arises from the composition of the vote message, containing $|00\rangle$, $|01\rangle$, and $|11\rangle$ states, with noise introducing the $|10\rangle$ state. Single qubits more easily transition to the $|10\rangle$ state, rendering the $|00\rangle$ and $|11\rangle$ states more susceptible. Lastly, on the IBM Quantum Computer 'ibm\_nairobi,' a 1.66\% noise state is observed, while on 'ibm\_lagos,' a 1.36\% noise state $|10\rangle$ is recorded, as depicted in Fig. \ref{fig:nairobi_lagos}. 

\clearpage

\section{Conclusion}
This paper introduces a straightforward voting protocol based on Amplitude Encoding, Blockchain, and Signature techniques. Despite its simplicity, the protocol offers a comprehensive set of essential properties, including binding, anonymity, non-reusability, verifiability, eligibility, and fairness in the voting process. To assess the protocol's performance in a real-world context for four candidates, we conducted simulations in a noisy environment, incorporating measurement errors and gate errors ranging from 0.1\% to 10\%. Furthermore, we executed the protocol on quantum hardware, specifically on IBM Quantum Machines—namely, ibm\_nairobi and ibm\_lagos—both of which are 7-qubit devices. Our findings indicate a minimal error rate of 1.66\% on ibm\_nairobi and 1.36\% on ibm\_lagos, showcasing the protocol's effectiveness and robustness.

\section*{Acknowledgement}
We recognize the utilization of IBM Quantum services for this work. We would like to acknowledge the IBM Quantum. The perspectives communicated are those of the authors and does not mirror the authority strategy or position of IBM or the IBM Quantum group. In this paper, we utilized ibmq qasm simulator which is a portion of the IBM Quantum Canary Processors.

\section*{Conflict of Interest}
The authors declare no competing interests.

\section*{Data Availability}
The complete code and documentation for this Qiskit-based Quantum Voting Protocol ––can be accessed on GitHub at the following link:
\href{https://github.com/illogicallylogical04/Qiskit-Implementation-of-Quantum-Voting-Protocol}{https://github.com/illogicallylogical04/Qiskit-Implementation-of-Quantum-Voting-Protocol}

\bibliographystyle{unsrt}  
\bibliography{references}  

\end{document}